\documentclass[aps,pre,twocolumn,float]{revtex4}
\usepackage{amsmath,bm,epsfig}
\usepackage{amssymb}

\DeclareMathAlphabet{\mathitbf}{OML}{cmm}{b}{it}

\newcommand{\sFrac}[2]{{\textstyle\frac{#1}{#2}}}

\newcommand{\xv}{\mathitbf x}

\newcommand{\uv}{\mathitbf u}
\newcommand{\fv}{\mathitbf f}

\newcommand{\calBold}[1]{\mbox{\boldmath${\cal #1}$}}
\newcommand{\mathBold}[1]{\mbox{\boldmath$#1$}}
\newcommand{\zv}{\mathitbf z}

\def\Fbox#1{\vskip1ex\hbox to 8.5cm{\hfil\fboxsep0.3cm\fbox{%
  \parbox{8.0cm}{#1}}\hfil}\vskip1ex\noindent}  

\newcommand{\B}[1]{{\bm{#1}}}
\let \= \equiv \let\*\cdot \let\~\widetilde \let\^\widehat \let\-\overline

\begin{document}
\title{Statistical Physics of Elasto-Plastic Steady States in Amorphous Solids: Finite Temperatures and Strain Rates}
\author{Smarajit Karmakar, Edan Lerner, Itamar Procaccia and Jacques Zylberg}
\affiliation{Department of Chemical Physics, The Weizmann
Institute of Science, Rehovot 76100, Israel}
\date{\today}
\begin{abstract}
The effect of finite temperature $T$ and finite strain rate $\dot\gamma$ on the statistical physics of plastic deformations in amorphous solids made of $N$ particles is investigated. We recognize three regimes of temperature where the statistics are qualitatively different. In the first regime the temperature is very low, $T<T_{\rm cross}(N)$, and the strain is quasi-static. In this regime the elasto-plastic steady state exhibits highly correlated plastic events whose statistics are characterized by anomalous exponents. In the second regime $T_{\rm cross}(N)<T<T_{\rm max}(\dot\gamma)$ the system-size dependence of the stress fluctuations becomes normal, but the variance depends on the strain rate. The physical mechanism of the cross-over is different for increasing temperature and increasing strain rate, since the plastic events are still dominated by the mechanical instabilities (seen as an eigenvalue of the Hessian matrix going to zero), and the effect of temperature is only to facilitate the transition. A third regime occurs above the second cross-over temperature $T_{\rm max}(\dot\gamma)$ where stress fluctuations become dominated by thermal noise. Throughout the paper we demonstrate that scaling concepts are highly relevant for the problem at hand, and finally we present a scaling theory that is able to collapse the data for all the values of temperatures and strain rates, providing us with a high degree of predictability.
\end{abstract}
\maketitle

\section{Introduction}

The simulational investigation of elasto-plastic steady states in amorphous solids concentrated in recent years on athermal and quasi-static conditions (AQS) \cite{04ML,06TLB,06ML,09MR,09LP}. The reasons for this are manifold: first, in athermal
conditions the separation between elastic regimes and plastic events is clear cut. Second, the mechanism for plastic
events is purely mechanical, and it can be understood entirely on the basis of the underlying inter-particle
potentials and dynamics \cite{10KLLP,10KLP}. Third, the simulations exhibit the existence of fascinating, highly correlated plastic events whose statistics abound with anomalous exponents that are not fully understood. On the other hand, current models of
elasto-plasticity tend to propose mean-field type theories in which plastic events are uncorrelated and are not
linked to mechanical instabilities, but are rather assumed to be triggered by purely thermal fluctuations \cite{79AK,79Arg,82AS,98FL,98Sol,07BLP}. If so, the interesting findings in the AQS regime might be entirely irrelevant for ``normal plasticity" at higher temperatures and strain rates. The aim of this paper is to examine this issue with some care. To this purpose we will explore here the effect of temperature and of strain rates, separately and together, to assess whether indeed the existence of finite temperatures and strain rates obliterate the relevance of the rich plethora of findings at the AQS conditions. Our conclusion is that this is far from being true \cite{10CLC,10HKLP}. In fact, even when temperature or strain rates are enough to make the stress and energy fluctuations `normal' (in a sense made precise below) the role of mechanical instabilities is still crucial. We thus need to reveal the mechanisms responsible for the cross-over in statistics and to propose a theoretical framework for the discussion of the statistical physics of elasto-plastic steady state taking into account all the essential ingredients. When done, as described below in this paper, we will own a description that unifies the statistical physics below and above the
crossovers from highly correlated to uncorrelated stress and energy fluctuations.

The structure of this paper is as follows. In Sect. \ref{numerics} we present the model glass former and the details of the numerical procedures employed throughout this study. In Sect. \ref{review} we present a short review of the essential results pertaining to elasto-plasticity in amorphous solids in AQS conditions. In Sect. \ref{thermal} we consider the effect of temperature on elasto-plasticity with very slow strain rates (where very slow means rates below the cross-over
to independent stress and energy fluctuations due to high strain rates). We show that the plastic events are assisted by thermal fluctuations but they are nevertheless dominated by mechanical instabilities. The cross-overs from highly correlated plastic events to independent fluctuations is considered in Sect. \ref{cross}, where the different mechanisms
for these cross-overs due to thermal fluctuations and due to high strain rates are treated separately. Throughout the paper we use scaling concepts to organize, compactify and exhibit the physics in its neatest form. This approach culminates in Sect. \ref{scale} where we
present the scaling function theory that unifies the statistical physics in the elasto-plastic steady state before and after the thermal-dominated and strain-rate-dominated cross overs. Sect. \ref{conclusions} offers a summary of the paper and concluding remarks.

\section{Model and Numerical Methods}
\label{numerics}
\subsection{System Definitions}
Below we employ a model glass-forming system with point particles of two `sizes' but of equal mass $m$
in two and three dimensions (2D and 3D respectively), interacting via a pairwise potential of the form
\begin{equation}\label{potential}
\phi\left(\!\frac{r_{ij}}{\lambda_{ij}}\!\right) =
\left\{ \begin{array}{ccl} \!\!\varepsilon\left[\left(\frac{\lambda_{ij}}{r_{ij}}\right)^{k} + \displaystyle{\sum_{\ell=0}^{q}}c_{2\ell} \left(\frac{r_{ij}}{\lambda_{ij}}\right)^{2\ell}\right] &\! , \! & \frac{r_{ij}}{\lambda_{ij}} \le x_c \\ 0 &\! , \! & \frac{r_{ij}}{\lambda_{ij}} > x_c \end{array} \right., \end{equation} where $r_{ij}$ is the distance between particle $i$ and $j$, $\varepsilon$ is the energy scale, and $x_c$ is the dimensionless length for which the potential will vanish continuously up to $q$ derivatives. The interaction lengthscale $\lambda_{ij}$ between any two particles $i$ and $j$ is $\lambda_{ij} = 1.0\lambda$, $\lambda_{ij} = 1.18\lambda$ and $\lambda_{ij} = 1.4\lambda$ for two `small' particles, one `large' and one `small' particle and two `large' particle respectively. The coefficients $c_{2\ell}$ are given by \begin{equation} c_{2\ell} = \frac{(-1)^{\ell+1}}{(2q-2\ell)!!(2\ell)!!}\frac{(k+2q)!!}{(k-2)!!(k+2\ell)}x_c^{-(k+2\ell)}.
\end{equation}
We chose the parameters
$x_c = 1.385$, $k=10$ and $q=2$.
The unit of length $\lambda$ is set to be the interaction length scale of two small particles, and
$\varepsilon$ is the unit of energy. Accordingly, the time is measured in units of $\tau_\star =\sqrt{m\lambda^2/\varepsilon}$.
The density for all systems is set to be $N/V = 0.85\lambda^{-2}$.
The glass transition temperature is $T_g \approx 0.46\varepsilon/k_B$, defined here by the condition $\tau_\alpha(T_g)=10^5\tau_\star$,
where $\tau_\alpha$ is the structural relaxation time.

\subsection{Methods}
\label{methods}
The work presented here is based on three types of simulational methods. The first type
corresponds to the athermal quasi-static (AQS)
limit $T \rightarrow 0$ and $\dot{\gamma} \rightarrow 0$,
where $\dot{\gamma}$ is the strain rate.
AQS simulations had been extensively analyzed
recently \cite{04ML,06TLB,06ML,07BSLJ,08TTLB,09LP} as a tool for investigating
plasticity in amorphous systems. In AQS simulations one starts from
a completely quenched configuration of the system, and applies an affine simple shear
transformation to each particle $i$ in our shear cell, according to
\begin{eqnarray}\label{affineTransformation}
r_{ix} & \rightarrow & r_{ix} + r_{iy}\delta\gamma\ , \nonumber\\
r_{iy} & \rightarrow & r_{iy} \ ,
\label{simpleShearTransformation}
\end{eqnarray}
in addition to imposing Lees-Edwards boundary conditions \cite{91AT},
and $\delta\gamma = \gamma - \gamma_0$ is a small strain
increment from some reference strain $\gamma_0$.
The strain increment $\delta\gamma$ plays a role analogous to the integration
step in standard MD simulations. We choose the basic strain increment step to be $\delta\gamma = 5\times 10^{-5}$ for all
system sizes simulated, and sample each plastic event using strain increments of at most
$2\times10^{-6}$ using the backtracking procedure described in \cite{09LP}.
The affine transformation (\ref{affineTransformation})
is then followed by the minimization of the potential
energy under the constraints imposed by the strain increment and the periodic
boundary conditions. We chose
the termination threshold of the minimizations to be $|\nabla_i U| = 10^{-9}$,
for every degree of freedom $x_i$.
Our method for locating saddle points is explained in Subsect. \ref{sdeb}.

The second simulation method employs the so-called SLLOD equations of motion
\cite{91AT}. For our constant strain rate 2D systems, they read
\begin{eqnarray}\label{sllod}
\dot{r}_{ix} & = & p_{ix}/m + \dot{\gamma}r_{iy}\ ,\nonumber \\
\dot{r}_{iy} & = & p_{iy}/m\ , \nonumber \\
\dot{p}_{ix} & = & f_{ix} - \dot{\gamma}p_{iy}\ , \nonumber \\
\dot{p}_{iy} & = & f_{iy}\ .
\end{eqnarray}
We use a leapfrog integration scheme for the above equations,
and keep the temperature constant by employing a modification of the Berendsen thermostat \cite{91AT},
measuring the instantaneous temperature with respect to a homogeneous shear flow.
The modification implies that we randomly re-partition the system into subsets of
about 500 particles, and utilize a set of Berendsen factors, with a different factor for each subset (instead of just one in the standard
algorithm). This modification was found necessary in order to reduce finite-size effects
due to sub-extensive statistics.
The integration time step was chosen to be $\delta t = 0.005$,
and we set the time scale for heat extraction at $\tau_T = 5.0\tau_\star$.

The third simulational method \cite{09LC} was employed to study systems at athermal conditions, but
at finite strain rates. This method utilizes the SLLOD equations of motion (\ref{sllod}), with
the addition of total momentum conserving damping forces, such that the force on the $i$'th particle is given by
\begin{equation}
\B f_i = \sum_{j\ne i} {\B f}_{ij} +
\frac{m}{\tau_T}\sum_{j \ne i}D_{ij}({\bf v_i} - {\bf v_j})
\end{equation}
where ${\B f}_{ij}$ is calculated from the pair potential $\phi_{ij}$ and $D_{ij}$ is given
as
\begin{equation}
D_{ij} = 1 - 2(r_{ij}/r_c)^4 + (r_{ij}/r_c)^8
\end{equation}
which vanishes smoothly at $r_{ij} = r_c$, where $r_c$ is the cutoff of the pair potential
$\phi_{ij}$. For integrating the equation of motion we will use a slightly modified version of the
Velocity-Verlet algorithm defined below ~\cite{97GW}
\begin{equation}
\begin{split}
\B r_i(t + \delta t)& = \B r_i(t) + \B v_i(t)\delta t + \sFrac{1}{2} \B f_{i}(t)(\delta t)^2 \ ,\\
\tilde{\B v}_i(t+\delta t) & = \B v_i(t) + \sFrac{1}{2} {\B f}_i(t) \ ,\\
{\B f}_i(t+\delta t) &=  {\B f}_i \left( {\B r}_i(t+\delta t), \tilde{\B v}_i(t+\delta t)\right)\ ,\\
{\B v}_i(t+\delta t) &= {\B v}_i(t) + \sFrac{1}{2}\left({\B f}_i(t)+
{\B f}_i(t+\delta t)\right)\delta t \ .
\end{split}
\end{equation}

Below we need to determine whether a thermal stressed system still resides in an original local minimum
of the athermal system or whether it had jumped to the basin of attraction of another local minimum.
This is done by taking a given configuration and minimizing its potential energy until it hits the
minimum. For a system that is stressed at a finite temperature one observes plastic events before the
mechanical instability threshold is reached. By minimizing the energy every 100 times steps in the simulation below we can identify such transitions by finding that the original minimum is no longer captured and a new
one replaced it.

For completeness we also carried out simulations in three dimensions, using the same binary mixture with the same interaction potential as in 2D but with a density $N/V=0.81$. Some result from these simulations are discussed below.

\section{Review of elasto-plasticity in AQS conditions}
\label{review}
We consider amorphous solids in the limit of zero temperature $T\to0$, subjected,
say, to shear deformation at vanishing low strain rates $\dot{\gamma}$,
with $\gamma$ our parametrization of the imposed deformation, see below.
An amorphous solid in the athermal limit must satisfy the following conditions \cite{06ML}: ({\it i}) the notion
of solidity requires that \emph{all} the eigenvalues of the Hessian matrix $\calBold{H}_{ij} \equiv
\frac{\partial ^2 U}{\partial \xv_j \partial \xv_i}$ are strictly positive.
({\it ii}) The amorphous nature of the considered systems is guaranteed by
demanding that the \emph{mismatch forces} $\mathBold{\Xi}_i \equiv \frac{\partial ^2 U}
{\partial \gamma \partial \xv_i}$ are non-zero and uncorrelated, i.e.
$\mathBold{\Xi}_i \ne 0, \langle \mathBold{\Xi}_i \mathBold{\Xi}_j \rangle \sim \delta_{ij}$.
({\it iii}) The limit $T\to 0$ implies that the system always resides in
a local minimum of the potential, with the forces
\begin{equation}\label{zeroForcesConstraint}
\fv_i = -\sFrac{\partial U}{\partial \xv_i} = 0\ ,
\end{equation}
at all times.

In AQS simulations the potential energy $U$
is a function of the imposed strain, parameterized by $\gamma$, and
of the particle coordinates $\xv_i(\gamma)$, $U = U(\{\xv_i(\gamma)\},\gamma)$.
Below we derive the explicit coordinate dependence on $\gamma$;
we consider deformations via parameterized
transformations on the particle coordinates $\mathBold{H}(\gamma) = \calBold{I} + \gamma\mathBold{h}$
(not to be confused with the Hessian $\calBold{H}$):
\begin{equation}\label{transformation}
\xv_i \to \mathBold{H}\cdot \xv_i + \uv_i \ ,
\end{equation}
where the non-affine coordinates $\uv_i$ are additional displacements
that assure that the zero-forces constraint (\ref{zeroForcesConstraint}) is fulfilled;
total derivatives with respect to strain in the athermal limit
should thus satisfy the zero-forces constraint. They
are carried out via
\begin{equation}
\frac{d}{d\gamma} = \frac{\partial }{\partial \gamma} + \frac{d \uv_i}{d \gamma}\cdot \frac{\partial}{\partial \uv_i}
=\frac{\partial }{\partial \gamma} + \frac{d \uv_i}{d \gamma}\cdot \frac{\partial}{\partial \xv_i} \ ,
\end{equation}
where the second equality results from from Eq.~(\ref{transformation}),
and here and below repeated indices are summed over.
The evolution of the non-affine coordinates $\uv_i$ can
be explicitly derived by requiring
that $ \frac{d \fv_i}{d \gamma}=0$:
\begin{equation}
\frac{d \fv_i}{d \gamma} =
\frac{\partial \fv_i}{\partial \gamma} + \frac{\partial \fv_i}{\partial \xv_j}\cdot
\frac{d \uv_j}{d \gamma} = 0\ .	\label{above}
\end{equation}
We refer to the full derivatives of the relaxation coordinates
with respect to strain as the non-affine velocities
$\mathBold{v}_i \equiv \frac{d \uv_i}{d\gamma}$. Inserting the definitions of the Hessian $\calBold{H}_{ij}$
and the mismatch forces $\mathBold{\Xi}_i$ in Eq. (\ref{above}), we obtain
\begin{equation}\label{foo1}
\mathBold{\Xi}_i + \calBold{H}_{ij}\cdot\mathBold{v}_j = 0\ .
\end{equation}
From here, the equation for the non-affine velocities $\{ \mathBold{v}_i \}$ is obtained by inverting (\ref{foo1}):
\begin{equation}\label{nonAffineVelocities}
\mathBold{v}_i = -\calBold{H}^{-1}_{ij}\cdot\mathBold{\Xi}_j\ .
\end{equation}
With an equation for the non-affine velocities $\mathBold{v}_i$, full derivatives with respect to
strain can be written as
\begin{equation}
\frac{d }{d\gamma} = \frac{\partial }{\partial \gamma} + \mathBold{v}_j\cdot \frac{\partial}{\partial \xv_j} \ ,
\end{equation}
and the equation of motion for the
coordinates $\xv_i(\gamma)$ becomes explicitly available:
\begin{equation}
\frac{d \xv_i}{d\gamma} = \mathBold{h}\cdot\xv_i + \mathBold{v}_i\ .
\end{equation}
\begin{figure}[!ht]
\centering
\includegraphics[scale = 0.35]{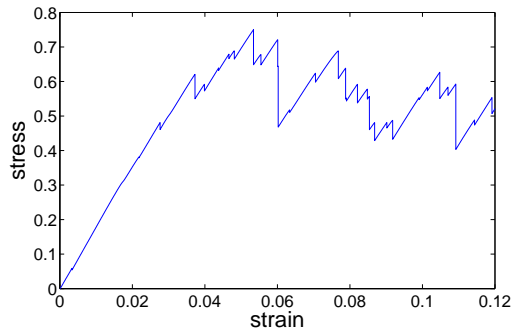}
\caption{A typical stress vs. strain curve in a system of 4096 particles under simple shear deformation
in two dimensions obtained in the athermal quasistatic limit.
Every elastic (reversible) increase in stress is followed by a sudden plastic (irreversible) drop in stress.}
\label{stress-strain}
\end{figure}

The above analysis of the AQS dynamics is valid as long as all the eigenvalues of the
Hessian are positive; this condition breaks down eventually upon increasing the external strain,
when reversible elastic branches are terminated by mechanical instabilities, as demonstrated in Fig.~\ref{stress-strain}.
These mechanical instabilities are associated with the vanishing of an eigenvalue of the
Hessian, when the local minimum at which the system resided develops an unstable direction
in coordinate space, which the system follows while undergoing an abrupt drop in energy. The
system then descends down the potential energy landscape until finding some other mechanicaly stable local minimum.
Associated with this drop in energy, denoted in the following by $\Delta U = U_{\rm initial} - U_{\rm final}$, is a change in stress,
denoted in the following by $\Delta \sigma = \sigma_{\rm initial} - \sigma_{\rm final}$; the change in stress is
not strictly constrained to be positive - in small systems one may encounter mechanical instabilities that result in an increase rather than a decrease of stress.
In this section we will provide a brief review of the physics of deformed systems approaching mechanical instabilities,
followed by a review of the statistics of the flow events in the steady flow state. We will demonstrate in the next section that the
mechanical instabilities remain highly relevant also in the second temperature regime $T_{\rm cross}<T<T_{\rm max}$.

\subsection{Mechanical instabilities}
As mentioned above, upon increasing the external strain, an eigenvalue of the Hessian eventually vanishes
at the inevitable onset of a mechanical instability. We denote the vanishing eigenvalue as $\lambda_P$, its
corresponding eigenfunction as $\bm{\psi}^{(P)}_i$ and the strain value at which
the upcoming instability occurs as $\gamma_P$, such that
\begin{equation}\label{foo4}
\lambda_P \to 0 \ \ \mbox{as} \ \ \gamma \to \gamma_P \ .
\end{equation}
Returning to Eq.~(\ref{nonAffineVelocities}), we
write the non-affine velocities in a normal mode decomposition \cite{04ML},
\begin{equation}
\mathBold{v}_i = \!-\calBold{H}_{ij}^{-1} \cdot\bm{\Xi}_j \!=
-\!\!\sum_k\frac{ \bm{\psi}^{(k)}_j\cdot \mathBold{\Xi}_j }
{\lambda_k}\bm{\psi}^{(k)}_i\ ,
\end{equation}
where $\lambda_k$ is an eigenvalue of the Hessian and $\bm{\psi}^{(k)}_j$ its corresponding eigenfunction:
$\calBold{H}_{ij}\cdot \bm{\psi}_j^{(k)} =\lambda_k \bm{\psi}_j^{(k)}$.
As $\lambda_P \to 0$, the above sum will be dominated by the diverging term, i.e.
\begin{equation}\label{foo2}
\mathBold{v}_i \to  -\frac{ \bm{\psi}^{(P)}_j\cdot \bm{\Xi}_j }{\lambda_P}\bm{\psi}^{(P)}_i \ \ \mbox{as}
\ \ \gamma \to \gamma_P\ .
\end{equation}
We now calculate derivatives of the potential energy with respect to strain in the vicinity of a mechanical
instability, say, at $\gamma_P$; when possible, we will only keep terms that consist of contractions
with the non-affine velocities $\bm{v}_i$, since these diverge as the system is strained towards $\gamma_P$,
cf. Eq.~(\ref{foo4}) and Eq.~(\ref{foo2}), hence they will dominate over any regular terms.
With $V=L^d$ denoting the volume of the system with linear size $L$, the stress $\sigma$ is defined via
\begin{equation}
V\sigma \equiv \frac{d U}{d \gamma} = \frac{\partial U}{\partial \gamma} + \bm{v}_j \cdot \frac{\partial U}{\partial \xv_j}
= \frac{\partial U}{\partial \gamma}\ ,
\end{equation}
where the second equality stems from the constraint of zero forces (\ref{zeroForcesConstraint}).
Since the non-affine velocities do not appear in the above expression for the stress, it is always regular,
even as $\gamma\to \gamma_P$. Taking another total derivative of the potential energy with respect to strain,
keeping only highest order terms in $\frac{1}{\lambda_P}$:
\begin{equation}\label{foo3}
\frac{d^2U}{d\gamma^2} \simeq \mathBold{v}_j\cdot
\frac{\partial ^2U}{\partial \xv_j \partial \gamma}	= \mathBold{v}_j\cdot \mathBold{\Xi}_j \simeq  -\frac{a}{\lambda_P} \ ,
\end{equation}
with $a=(\bm{\psi}^{(P)}_j\cdot \bm{\Xi}_j)^2$.
Taking a last total derivative of the potential energy with respect to strain
requires an expression for $\frac{d \bm{v}_i}{d \gamma}$; with the notation
$\calBold{T}_{ijk} \equiv \frac{\partial ^3U}{\partial \xv_k\partial \xv_j\partial \xv_i}$ ,
an expression can be obtained by writing
\begin{eqnarray}
-\frac{d^2 \fv_i}{d \gamma^2} & = & \frac{d}{d\gamma}\left(\bm{\Xi}_i
+ \calBold{H}_{ij}\cdot\bm{v}_j\right) \nonumber \\
& = &  \frac{\partial \bm{\Xi}_i}{\partial \gamma} + \calBold{T}_{ijk}: \bm{v}_j\bm{v}_k  \\
&& + 2\frac{\partial \calBold{H}_{ij}}{\partial \gamma} \cdot \bm{v}_j
+ \calBold{H}_{ij}\cdot\frac{d \bm{v}_j }{d\gamma} = 0\ , \nonumber
\end{eqnarray}
and inverting for $\frac{d \bm{v}_i}{d \gamma}$.
Keeping the highest order terms in $\frac{1}{\lambda_P}$, we find that in the vicinity of $\gamma_P$,
\begin{equation}
\frac{d \bm{v}_i}{d \gamma} \simeq -\calBold{H}^{-1}_{ij}\cdot \left(\calBold{T}_{jk\ell}:\bm{v}_k \bm{v}_\ell\right)\ ,
\end{equation}
thus
\begin{equation}\label{foo5}
\frac{d^3U}{d \gamma^3}\simeq \frac{d \bm{v}_j}{d \gamma}\cdot \bm{\Xi}_j
\simeq \calBold{T}_{ijk}\vdots\bm{v}_i\bm{v}_j\bm{v}_k \simeq -\frac{b}{\lambda_P^3}\ ,
\end{equation}
with $b=(\bm{\psi}^{(P)}_\ell\cdot \bm{\Xi}_\ell)^3\,
\calBold{T}_{ijk}\vdots\bm{\psi}^{(P)}_i\bm{\psi}^{(P)}_j\bm{\psi}^{(P)}_k$.
Combining results (\ref{foo3}) and (\ref{foo5}), we arrive at the differential equation
\begin{equation}
\frac{d}{d\gamma}\left(\frac{1}{\lambda_P}\right) \sim \frac{1}{\lambda_P^3}\ .
\end{equation}
The solution of the above differetial equation, together with the boundary condition $\lambda_P(\gamma_P) = 0$, is \cite{10KLLP},
\begin{equation}
\lambda_P \sim \sqrt{\gamma_P - \gamma}\ . \label{lamtozero}
\end{equation}
We should emphasize at this point that the potential energy should, in principle, be written as a series expansion in terms of components of the strain tensor $\calBold{\epsilon} \equiv \frac{1}{2}(\mathBold{H}^T\cdot\mathBold{H} - \calBold{I})$ \cite{10KLP}.
Then, additional terms may appear in potential energy derivatives, given a parametrization of $\mathBold{H}(\gamma)$.
Here, for the sake of simplicity, we directly utilize a parameterized notion of deformation via the parameter $\gamma$.
Since we were only interested in singular terms near mechanical instabilities, our results are equally valid.

\subsection{Flow events statistics}
\begin{figure}[!ht]
\centering
\includegraphics[scale = 0.50]{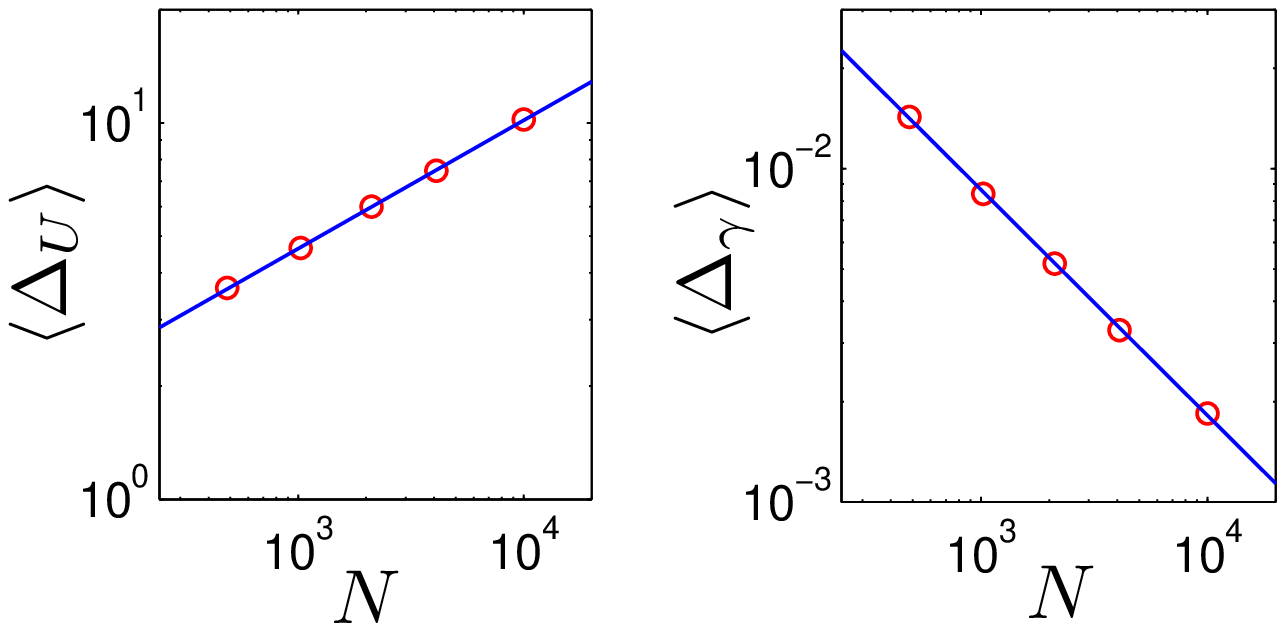}
\includegraphics[scale = 0.50]{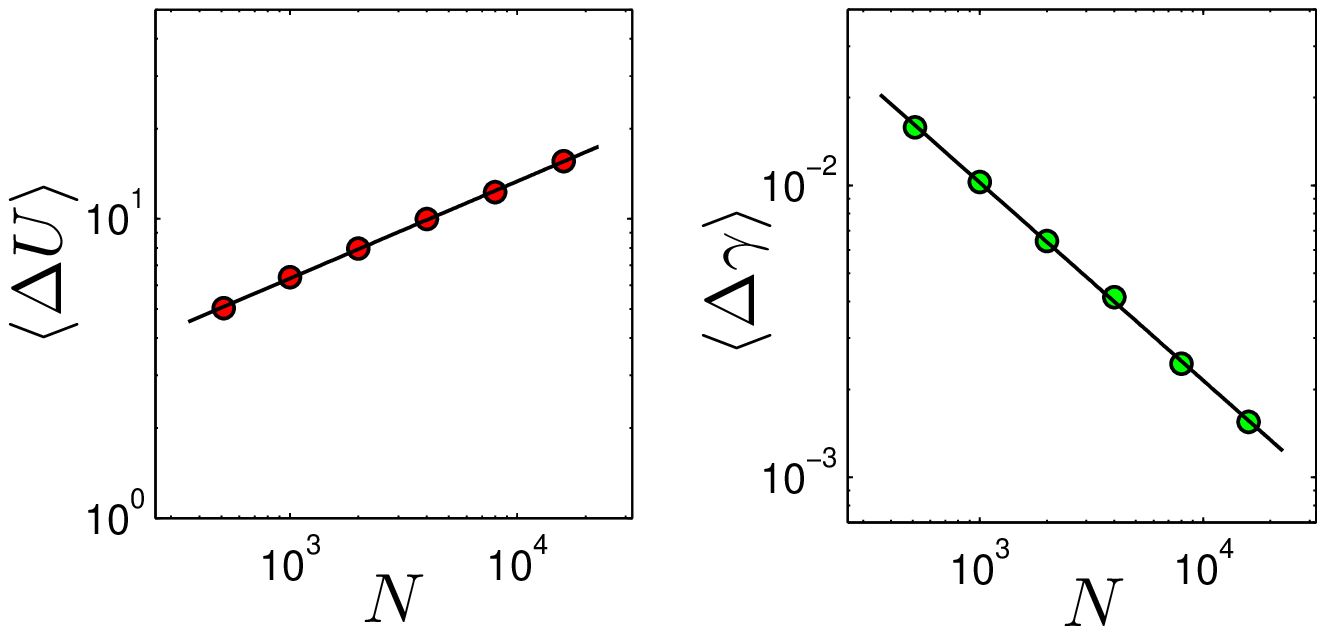}
\caption{ Mean energy drop $\langle \Delta U \rangle$ (left panels) and mean strain interval
$\langle \Delta \gamma \rangle$ (right panels) as functions
of system size, measured in AQS simulations of steady plastic flow of the
system described in Sect.~\ref{numerics}. Upper panels: two dimensions. Lower panels: three dimensions. The continuous lines represent the scaling laws (\ref{mean1}) and (\ref{mean2}). The scaling exponents are the same in 2D and 3D.}
\label{meanScaling}
\end{figure}

When subjecting an amorphous solids to external shear strain,
it tends to set up an elasto-plastic steady flow state after a short transient of a few percent strain.
In the steady flow state, the statistics of the energy drops $\Delta U$, the stress drops $\Delta \sigma$ and
the strain intervals between successive flow events $\Delta \gamma$ become stationary.
In particular, one finds that the averages of these quantities obey the following scaling relations
\begin{eqnarray}\label{mean1}
&&\langle \Delta U \rangle \sim \bar\epsilon N^\alpha\ , \quad \langle \Delta \sigma \rangle \sim \bar s N^\beta\\
&&\langle \Delta \gamma \rangle \sim N^\beta\ . \label{mean2}
\end{eqnarray}
In Fig.~\ref{meanScaling} the mean energy drop $\langle \Delta U \rangle$
and mean strain interval $\langle \Delta \gamma \rangle$ for our model system
are displayed, together with the scaling laws (\ref{mean1}). In the upper panels we show results in two dimensions
and in the lower panel in three dimensions,
In the present model we find that $\alpha \approx 1/3$ and $\beta \approx -2/3$ in both 2D and 3D.

A scaling relation $\alpha -\beta=1$ follows from the average
energy balance equation, cf. [5]
\begin{equation}
\sigma_Y \langle \Delta \sigma \rangle V/\mu = \langle \Delta U \rangle\ , \label{bal}
\end{equation}
where $\sigma_Y$ is the yield stress (the mean stress in the steady state flow state, in the AQS limit),
and $\mu$ is the shear modulus.

\subsection{The scaling of the variance of stress fluctuations}

\begin{figure}[!ht]
\centering
\includegraphics[scale = 0.42]{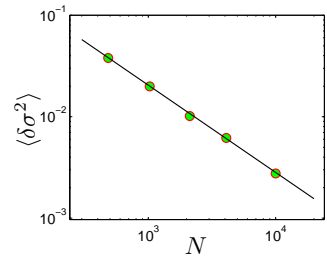}
\caption{ Stress fluctuations $\widetilde{\langle \delta \sigma ^2} \rangle$ of the steady flow state
of the studied model. The continuous line represents the scaling law (\ref{stressFlucsScaling}).}
\label{aqsStressFluctuations}
\end{figure}

For our purpose of distinguishing clearly between different regimes of elasto-plastic statistical physics
it is advantageous to measure the properties of stress fluctuations, and in particular of the variance.
This quantity will reflect the change in physics in the different regimes. In AQS conditions
the variance of stress fluctuations $\widetilde{\langle \delta \sigma ^2} \rangle \equiv \langle (\sigma - \sigma_Y)^2 \rangle$
also exhibits anomalous scaling laws with the system size. In particular we find
\begin{equation}\label{stressFlucsScaling}
\widetilde{\langle \delta \sigma ^2} \rangle  \sim N^{2\theta} \ ,
\end{equation}
with $\theta \approx -0.4$ in the studied model, cf. Fig~\ref{aqsStressFluctuations}.
One should notice the difference between the exponent characterizing the $N$
dependence of $\sqrt{\widetilde{\langle \delta \sigma ^2\rangle}}$ and of the athermal mean
stress drop $\langle \Delta \sigma\rangle$, in the sense that $\theta\ne \beta$. This difference is
due to very strong correlations between elastic increases and plastic drops.

\section{The Mechanism of Thermally Activated Plasticity}
\label{thermal}

As explained above, in our thinking about thermal effects on elasto-plasticity in finite systems strained at finite strain rates and finite
temperatures there exist three temperature regimes in which the
dynamics and the statistics of plastic events are qualitatively different. In this section we investigate the
second regime $T_{\rm cross}(N) \le T \le T_{\rm max}(\dot\gamma)$ where $T_{\rm cross}(N)$ is system size dependent as explained in the following subsection.

\subsection{The first crossover in statistics}

The first cross-over in statistics from avalanches to independent statistics occurs at the temperatures $T\approx T_{\rm cross}$ where the cross-over temperature is
estimated as follows \cite{10HKLP}: during a plastic drop the energy released spreads out quickly in the system
on the time scale of elastic waves. Thus every particle shares an energy of the order of $\bar \epsilon N^\alpha/N=\bar \epsilon N^\beta$. On the other hand the typical scale of thermal energy per particle is $ T $ (in units of Boltzmann's constant). We thus expect thermal effects to start overwhelming the statistics of athermal plastic events when
\begin{equation}
\bar \epsilon N^\beta \sim  T_{\rm cross}  \ . \label{crossT}
\end{equation}
This equality will hold when the system size $L=\xi_2$, where $(\xi_2/\lambda)^d = N$. Substituting the last
equality in Eq. (\ref{crossT}) and then solving for $\xi_2$ we find
\begin{equation}
{\xi_2}/{ \lambda } = \left[{T_{\rm cross}}/{\bar\epsilon}\right]^{1/d\beta} \ . \label{xi2}
\end{equation}
Note that Eq. (\ref{crossT}) implies (since $\beta<0$) that the change in statistics occurs at a temperature that decreases when the system size increases, meanings
that AQS statistics will pertain only for small systems. We also understand the meaning of the cross-over
due to thermal effects: the thermal agitation reaches just the necessary level to compete with the stored
elastic energy per particle. The avalanches are made of the primary plastic
instability triggering all the other, close to instability regions, to flip in tandem and relax. These ripe regions
which are sufficiently close to instability to react to the primary instability, are all destroyed by the
thermal fluctuations such that the primary instability remains naked, turning the statistics of the plastic events
from anomalous to normal.

\begin{figure}
\centering
\hskip -1.0 cm
\includegraphics[scale = 0.45]{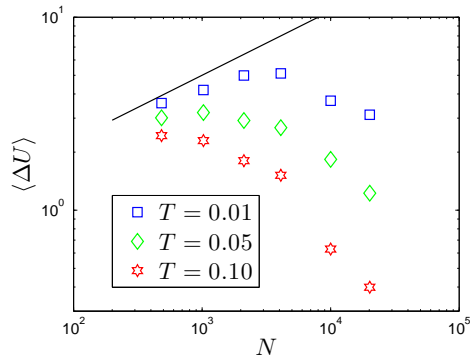}
\caption{The mean energy drop in plastic events as a function of system size for different temperatures, see inset.
The straight line represents the AQS scaling of mean energy drops.}
\label{mechanism}
\end{figure}

This picture can be demonstrated directly by measuring the energy drop in a plastic event in quasi-static
conditions but at different temperatures. The measurement is done by stopping the thermal molecular dynamics
simulation, followed by quenching the system to a very low temperature of $T=10^{-3}$ on a time scale of 100$\tau_\star$. This procedure allows for the completion of any plastic activity. Then the potential energy is minimized and the AQS scheme is employed to measure the energy drop in the first upcoming mechanical instability. This method allows us to probe directly the consequences of the thermal agitation on the ability of the system to undergo an avalanche. The results of such measurements are shown in Fig. \ref{mechanism}.
We see that the potential AQS energy drops are sapped out by the thermal agitation. Only at the lowest temperature of $T=0.01$ the energy drops approach the AQS limit for small systems, but even for this low temperature larger systems
cannot come close to the AQS limit. This is all in accordance with the estimates in Eqs. (\ref{crossT}) and (\ref{xi2}). We stress that this cross-over occurs at quasistatic conditions (but not only) and has nothing to do with $\dot\gamma$.

\begin{figure}
\centering
\hskip -1.0 cm
\includegraphics[scale = 0.45]{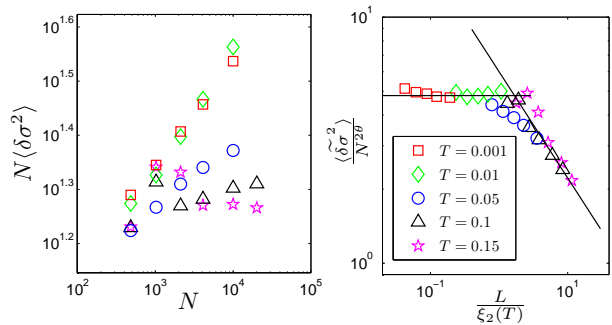}
\caption{Left panel: the variance of the stress fluctuations (multiplied by $N$) as a function of the system size $N$ for a 2D system and for various temperatures, at a strain rate $\dot\gamma=2\times 10^{-6}$. Right panel:
The scaling function $g(x)$, cf. Eq. (\ref{fun}) for the data in the left panel. Note the cross-over for $x$ of the order of unity as predicted by Eq. (\ref{xi2ofT}). The power law decrease at low values of $x$ are in agreement with the prediction of $\zeta\approx 0.33$. The two black lines represent the theoretical prediction for the scaling function $g(x)$ for $x\ll 1$ and
for $x\gg 1$. Note that the full scaling function depends on $\dot\gamma$, and the present one is an approximate
version for $\dot\gamma\to 0$.}
\label{gofx}
\end{figure}

An additional direct demonstration of the first thermal cross-over is obtained by measuring the stress fluctuations. Recall that at AQS conditions these fluctuation exhibit anomalous scaling , cf. Fig. \ref{aqsStressFluctuations}. At higher temperatures the data in Fig. \ref{gofx} indicate a clear cross-over to independent stress fluctuations in which $\langle \delta \sigma^2\rangle \sim N^{-1}$.

To capture the temperature and size dependence of the variance, and to demonstrate unequivocally the thermal cross-over, we first need to separate the thermal (vibrational) contribution from the mechanical contributions to $\langle \delta \sigma^2\rangle$. We write
\begin{equation}
\langle\delta\sigma^2\rangle = \langle\delta\sigma^2\rangle_T +
\widetilde{\langle\delta\sigma^2\rangle}\ , \label{sigandT}
\end{equation}
where $ \langle\delta\sigma^2\rangle_T$ denotes the thermal contribution which can be read from
Eq. (10) of Ref. \cite{07IMPS}, i.e.
\begin{equation}
 \langle\delta\sigma^2\rangle_T \approx \mu T/V \ .
\end{equation}
For the mechanical part we introduce a scaling function which exhibits the thermal cross-over. In other words, we propose a scaling function $g_2(x)$ to describe
the system-size and temperature dependence of the mechanical part of the variance. To this aim we
introduce the temperature dependent natural scale $\xi_2(T)$,
\begin{equation}
\xi_2(T) \equiv  \lambda \left[T/{\bar\epsilon}\right]^{1/d\beta} \ . \label{xi2ofT}
\end{equation}
The scaling function $g_2(x)$ is a function of the dimensionless ratio $L/\xi_2$:
\begin{equation}
\widetilde{\langle\delta\sigma^2\rangle}(N,T) =  \bar s^2N^{2\theta} g_2(L/\xi_2(T)) \ .
\end{equation}
The dimensionless scaling function $g_2(x)$ must satisfy
\begin{eqnarray}
g_2(x)&\to& g_0;  ~{\rm for}~x\to 0 \ , \nonumber\\
g_2(x)&\to& g_0 x^\zeta ~{\rm for  }~ x\to \infty \ . \label{fun}
\end{eqnarray}
The first of these requirements means that the fluctuation are in accordance with the athermal limit. The
second means that after the cross-over the fluctuations of the stress become intensive,
requiring $\zeta=-d(1+2\theta)$. We compute $\zeta\approx -0.4$ in 2D. Note that in Ref. \cite{10HKLP}
the same scaling function was written in terms of $1/x$ instead of $x$.

We present tests of the scaling function in  Fig. \ref{gofx}.
Examining the right panel of Fig. \ref{gofx} we see that the thermal cross-over is demonstrated very well where expected, i.e. at values of $x$ of the order of unity. The asymptotic behavior of the scaling functions agrees satisfactorily with the theoretical prediction which are indicated by the black lines.

\subsection{Strain dependent energy barriers}
\label{sdeb}
The aim of this section is to argue that for temperatures that are not too high the plastic events under
external strain are dominated by mechanical instabilities and are only assisted by thermal fluctuations.
It is crucial at this point to clarify what we mean by `not too high'. We will argue that in the elasto-plastic steady state there exists a `typical barrier for thermal activation'. Denoting this typical barrier by $\Delta G_{\rm typ}$
we estimate $T_{\rm max}$ by comparing the typical escape time over the typical barrier
$\tau\equiv \tau_\star\exp{(\Delta G_{\rm typ}/T)}$ to $\sigma_Y/(\mu\dot\gamma)$. The ratio $\tau\mu\dot\gamma/\sigma_Y$ measures the effect of the strain rate and when it exceeds some threshold, $T_{\rm max}$ should become a decreasing
function of $\dot\gamma$. We estimate $\Delta G_{\rm typ}$ below and address the effect of strain rate in more detail in the next two Sections. The present considerations apply for the temperature range $T_{\rm cross} \le T \le T_{\rm max}$.

Consider an athermal amorphous solid close to a
mechanical instability at $\gamma_P$; there, an eigenvalue $\lambda_P$ of the
Hessian vanishes, as seen in Eq. (\ref{lamtozero}), $\lambda_P \sim \sqrt{\gamma_P - \gamma}$.
Denoting $\calBold{\psi}^{(P)}$ the eigenvector associated with
the vanishing eigenvalue $\lambda_P$, we define the reaction coordinate $\tilde s$ as
the displacement of the system from the minimum of $U$ at $\xv^m$ in
the direction of $\calBold{\psi}^{(P)}$, i.e.
\begin{equation}
\xv_i = \xv^m_i + \tilde s \calBold{\psi}^{(P)}_i\ .
\end{equation}
We expand the potential up to third order in $\tilde s$:
\begin{equation}\label{expansionNearGP}
U(s) \simeq U(\xv^m) + \sFrac{1}{2}\lambda_P \tilde s^2 + \sFrac{1}{6}{\cal T}_P \tilde s^3
+ {\cal O}(\tilde s^4)\ ,
\end{equation}
where ${\cal T}_P \equiv \frac{\partial ^3 U}{\partial \tilde s^3} =
\frac{\partial ^3 U}{\partial \xv_k\partial \xv_j\partial \xv_i}\vdots
\calBold{\psi}^{(P)}_i \calBold{\psi}^{(P)}_j \calBold{\psi}^{(P)}_k$.
The relation (\ref{expansionNearGP})
is valid for $\gamma\to\gamma_P$; this form of the potential energy
results in the existence of a saddle point at $\tilde s=-\sFrac{2\lambda_P}{{\cal T}_P}$, which
translates to the real-space position of the saddle point at
\begin{equation}
\xv_i^s \equiv \xv^m_i - \sFrac{2\lambda_P}{{\cal T}_P}\calBold{\psi}^{(P)}_i\ .
\end{equation}
We define the energy barrier $\Delta E(\gamma)$ as the difference between the
potential at the saddle point $U(\xv^s)$, and the potential at the minimum $U(\xv^m)$,
i.e.
\begin{equation}\label{barrierDefinition}
\Delta E(\gamma) = U(\xv^s) - U(\xv^m)\ .
\end{equation}
It was shown in Ref. \cite{06MLa} that
the energy barrier $\Delta E(\gamma)$ obeys the scaling law
\begin{equation}\label{energyBarrierScaling}
\Delta E(\gamma) = c(\gamma_P - \gamma)^\frac{3}{2}\ ,
\end{equation}
over very large strain intervals of up to $10^{-2}$. Indeed, solving
$\frac{dU}{d\tilde s}=0$ in Eq.~(\ref{expansionNearGP}) results in the
estimation of the energy barrier of
$\Delta E(\gamma) \sim \frac{2\lambda_P^3}{3{\cal T}_P^2}$.
Plugging Eq. (\ref{lamtozero})
into this relation, assuming (as is the case in a saddle node bifurcation) that ${\cal T}_P$ is not singular near $\gamma_P$, we obtain
\begin{equation}
\Delta E(\gamma) \sim \frac{2\lambda_P^3}{3{\cal T}_P^2} \sim (\gamma_P - \gamma)^\frac{3}{2}\ .
\end{equation}
Note that the scaling $\lambda_P\sim\sqrt{\gamma_P-\gamma}$ holds only extremely close to
$\gamma_P$ (only vanishingly close to $\gamma_P$ as $N\to\infty$),
while the scaling of the energy barrier, Eq.~(\ref{energyBarrierScaling}),
holds along strain scales that are larger by orders of magnitude, \cite{06MLa}.
We show below that the range of strain on which the scaling law (\ref{energyBarrierScaling}) holds,
as well as the pre-factor $c$ in (\ref{energyBarrierScaling}) are
independent of system size.

To this aim we choose another reaction coordinate denoted $s$ as the
displacement of the coordinates $\xv$ from the minimum at $\xv^m$,
but this time directed towards the saddle point, i.e.
\begin{equation}
\xv_i = \xv^m_i + s \hat{\zv}_i\ ,
\end{equation}
where
\begin{equation}\label{saddleDirection}
\hat{\zv}_i \equiv \frac{\xv^s_i - \xv^m_i}{|\xv^s - \xv^m|}\ .
\end{equation}
\begin{figure}[ht]
\centering
\includegraphics[scale = 0.55]{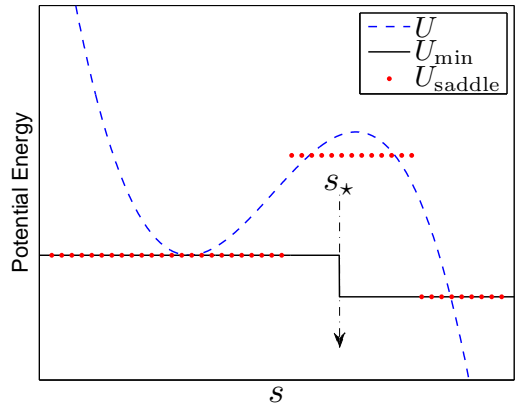}
\caption{Illustration of the procedure of finding the potential barriers.
Since the exact direction of the barrier $\hat{\zv}$ is unknown, the initial guess
brings the system to a state in which $U > U_{\rm saddle}$.
The point $s_\star$ is the displacement along the estimated $\hat{\zv}$ at which
the system leaves the original basin of attraction. At $s_\star$ the saddle point $\xv^s$
can be detected by minimizing the square of the gradient function $| \nabla_i U |^2$. }
\label{cartoon}
\end{figure}
Since it is not possible to know the direction $\hat{\zv}(\gamma)$ far away from $\gamma_P$, we
begin our measurements of $\Delta E$, for various system sizes, very close to $\gamma_P$, where
$\hat{\zv}(\gamma \to \gamma_P) \to \calBold{\psi}^{(P)}$, and set $\hat{\zv} = \calBold{\psi}^{(P)}$ as the initial guess for $\hat{\zv}$.
Then, to find the energy barrier, we displace the system along $\hat{\zv}$ by small increments of the reaction coordinate $s$,
and minimize the potential energy after each displacement.
For small displacements the minimization brings the system back to $\xv^m$; however, at some displacement $s_\star$,
the system does not return to the minimum at $\xv^m$, but rather
finds a different minimum, see Fig.~\ref{cartoon}. At the displacement value $s_\star$ at which
the system leaves the original basin of attraction, we minimize the square of the gradient function,
$| \nabla_i U |^2$. This brings us to the required saddle point $\xv^s$.
After finding the location of the saddle point $\xv^s$, we calculate the exact direction of $\hat{\zv}$
according to Eq.~(\ref{saddleDirection}), and the energy barrier according to Eq.~(\ref{barrierDefinition}).
The direction $\hat{\zv}$ is then recorded and used as the initial guess for the next iteration
in which the strain is further decreased.

\begin{figure}[ht]
\centering
\includegraphics[scale = 0.55]{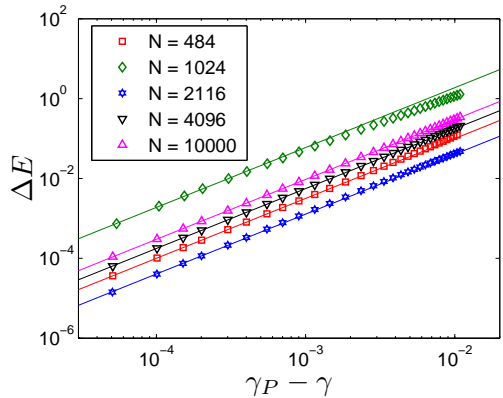}
\caption{Scaling of energy barriers for various system sizes. The slope of the continuous lines is 3/2.}
\label{energyBarrierScalingFig}
\end{figure}

In Fig.~\ref{energyBarrierScalingFig}, we present typical $\Delta E(\gamma)$ vs. $\gamma_P - \gamma$ for
systems of size $N=484,1024,2116,4096$ and $10000$. Clearly, as mentioned above, both
the pre-factor and the range in strain for which the scaling law (\ref{energyBarrierScaling}) holds
are independent of $N$.

Mechanical instabilities always involve the vanishing of the lowest eigenvalue of the Hessian;
this observation, together with the above findings, imply that the process of loosing mechanical
stability is reflected in the lowest eigenvalue of the Hessian only in some interval $\gamma_P - \gamma$
which is indeed system-size dependent. However, the same process is initialized at strain intervals
that are system size independent, of the order of tenths of a percent in strain, see Fig.~\ref{energyBarrierScalingFig}.
This, in turn, implies that the strain-induced reduction of energy barriers is highly relevant for the discussion of
thermally activated plasticity, for any system size.

\subsection{Thermal Activation of Plastic Events}

\begin{figure}
\centering
\includegraphics[scale = 0.45]{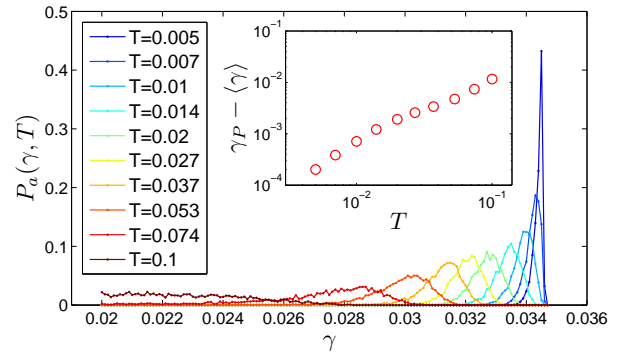}
\caption{The probability for thermally induced plastic event as a function of $\gamma$ for different
temperatures increasing from right to left. The system is of size $N=484$ and it has a mechanical instability
at $\gamma_P=0.0346$. Inset: The mean distance $\gamma_P-\langle\gamma\rangle$ of the distributions for different temperatures. Here $\dot\gamma=2\times 10^{-6}$.}
\label{activated}
\end{figure}

At higher temperatures this barrier can be overcome when $\gamma<\gamma_P$. To see the effect of temperature
explicitly one introduces the probability to undergo a thermally activated plastic event at the strain value $\gamma$, denoted as $P_a(\gamma,T)$ \cite{10CLC}. This probability was measured for a range of temperatures and is displayed in Fig.~\ref{activated} for a system with $N=484$ and an instability at $\gamma_P=0.0346$. This probability was measured at $\dot\gamma=2\times 10^{-6}$ by simulating a single elastic
branch with randomized initial velocities corresponding to a given temperature. For each member of the ensemble
we detect the value of the strain $\gamma$ at which the system leaves the basin of attraction of the athermal local minimum.
For low temperatures the mean of the distribution is very close to $\gamma_P$ and the distribution is sharp.
As temperature is increased the distributions move to the left, allowing a transition at lower values of $\gamma$,
with $\gamma_P-\langle \gamma\rangle$ being an increasing function of $T$. At about $T=0.1$ for this model
the distribution flattens out, and for higher temperatures thermal noise is dominant over the mechanical
instability. Using for this system $\sigma_Y\approx 0.5$, $\mu\approx 15$ and $\tau_\star\approx 1$ can estimate roughly $\Delta G_{\rm typ}\approx 1$.

\subsection{The third regime $T>T_{\rm max}$}

Having found an estimate for $\Delta G_{\rm typ}$ at the present value of the strain rate, we can also estimate $T_{\rm max}(\dot\gamma)$ as
\begin{equation}
T_{\rm max}\approx \frac{\Delta G_{\rm typ}}{\log(\sigma_Y/\mu\dot\gamma \tau_\star)} \ .
\end{equation}
Note that as a function of $\dot\gamma$ this value changes appreciably. For the present system this means that at temperatures higher than about 0.1 the
plastic events are dominated by thermal noise. Since the glass transition in this model is estimated
to occur about about $T_g \approx 0.46$, we expect to have a sizeable region $T_{\rm max} < T < T_g$ where
a theory assuming that plastic events are uncorrelated and dominated by thermal noise might be a good
model of the actual physics.

\section{Chopping Off the Avalanches: Strain Rate Effects}
\label{cross}

In this section we explain that increasing the strain rate results again (as for increasing the temperature) in turning the AQS anomalous
stress fluctuation to a normal process, but the physical mechanism is very different. Here the destruction of
the correlated events is due to the forcing (by the faster strain rate) of simultaneous plastic drops, not letting
them enough time to be correlated. To see this mechanism with clarity we need to expose a second length scale in addition
to $\xi_2(T)$ that was defined in Eq. (\ref{xi2ofT}).  This second length has to do with the strain rate $\dot\gamma$.

\subsection{The typical length associated with strain rate}
 We start by substituting Eq. (\ref{mean1}) in Eq. (\ref{bal}) to obtain the scale $\bar s$ \cite{10HKLP}. With $\lambda $ being the unit of length we write:
\begin{equation}
\bar s= \frac{\bar \epsilon\mu}{\sigma_Y \lambda^d} =\frac{\bar\epsilon\mu\rho}{\sigma_Y m} \ .
\end{equation}
Consider next the rate at which work is being done at the system and balance it by the energy dissipation in the steady state,
\begin{equation}
\sigma_Y \dot \gamma V =\langle \Delta U \rangle/\tau_{\rm pl} \ , \label{work}
\end{equation}
where $\tau_{\rm pl}$ is the average time between plastic flow events. This time is estimated as the elastic rise time which is
\begin{equation}
\tau_{\rm pl} \sim \frac{\langle\Delta \sigma \rangle}{\mu\dot\gamma} \sim \frac{\bar\epsilon N^\beta}{\sigma_Y \lambda^d \dot\gamma} \ .
\end{equation}
Next we note that $\tau_{\rm pl}$ decreases when $N$ increases. On the other hand there exists another crucial time scale in the system, which is the elastic relaxation time
$\tau_{\rm el} \sim L/c$,
 where $c$ is the speed of sound $c=\sqrt{\mu/\rho}$. Obviously this time scale {\em increases} with $N$ like $N^{1/d}$. There will be therefore a typical scale $\xi_1$ such that for a system of scale $L=\xi_1$ these times cross. At that size the system cannot equilibrate its elastic energy before another plastic event is triggered, and multiple avalanches must be occurring simultaneously in different parts of the system, each of which has a bounded magnitude.
We estimate $\xi_1$ from $\tau_{\rm el} \sim \tau_{\rm pl}$, finding
\begin{equation}
(\xi_1/c) \sim \frac{\epsilon [N(\xi_1)]^\beta}{\sigma_Y \lambda ^d\dot\gamma}\sim \frac{\epsilon [\xi/\lambda]^{d\beta}}{\sigma_Y  \lambda^d\dot\gamma} \ .
\end{equation}
Using now the obvious fact that $N(\xi_1) \sim (\xi_1/\langle \lambda \rangle)^d$ we compute
\begin{equation}
\frac{\xi_1}{\lambda } \sim \left[\left(\frac{\bar\epsilon}{\sigma_Y  \lambda ^d}\right)\, \left(\frac{c}{ \lambda \dot\gamma}\right) \right]^{1/(1-\beta d)} \ . \label{xi1}
\end{equation}
We observe the singularity for quasi-static strain when $\dot \gamma\to 0$, where $\xi_1$ tends to infinity, in agreement with the results of quasi-static calculations. At low temperatures, before the thermal energy scale becomes important, the size of plastic flow events can be huge indeed. Note that there exists a difference between our estimate of $\xi_1$ and that of Ref.~\cite{09LC}.

\subsection{The effect of simultaneous plastic events: the Herschel-Bulkley law}
\begin{figure}
\centering
\includegraphics[scale = 0.40]{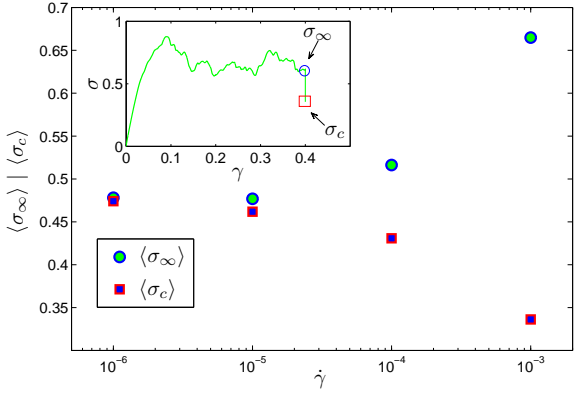}
\caption{The mean flow stress $\langle \sigma_\infty \rangle$ and the mean configurational stress $\langle \sigma_c\rangle$ as a function
of strain rate for a system with $N=1024$. Inset: a typical drop in the strain when the system is minimized to the nearest local minimum. }
\label{HBf}
\end{figure}

Due to the faster strain rates plastic events do not have time to cooperate and provide us with correlated avalanches. To exemplify the natural emergence of the length scale $\xi_1$ of Eq. (\ref{xi1}) in this context, we measure the difference between the mean flow stress measured at the steady state (denoted $\langle \sigma_\infty\rangle$) and the configurational stress of the corresponding local minimum (denoted $\langle \sigma_c\rangle$), see inset in Fig. (\ref{HBf}). We expect this difference to go to zero in AQS conditions (since all the plastic events are discharged during the avalanches) and to increase
with the strain rate. This difference
is found by stopping an athermal simulation run at a given strain rate $\dot \gamma$ (cf. Subsect. \ref{methods})
and quenching the system to hit the local minimum where the configurational stress is measured. The quenching is achieved first by running athermal molecular dynamics without increasing the strain for 100$\tau_\star$ to allow for any
plastic activity to complete, followed by a potential energy minimization. This procedure results in a stress drop
as seen in the inset of the right panel of Fig. \ref{HBf}. The amount of stress drop is determined by
the number of plastic events that occur starting at the point in time when the strain increase was stopped.

Plotting individually $\langle \sigma_\infty\rangle$ and $\langle \sigma_c \rangle$ for one system size as a function
of strain rate we note the relation of the present measurement to the Herschel-Bulkley law \cite{26HB} which relates
the mean flow stress to the strain rate,
\begin{equation}
\langle \sigma_\infty \rangle =\sigma_Y+ \hat s \dot\gamma^\chi \ . \label{HBL}
\end{equation}
Here $\sigma_Y$ is the mean flow stress in the AQS limit and $\chi$ is an exponent. This law is supposed to be $N$ independent
at large values of $N$.
The power-law dependence implied by Eq. (\ref{HBL}) is clearly seen in Fig.~\ref{HBf}.

\begin{figure}
\centering
\includegraphics[scale = 0.40]{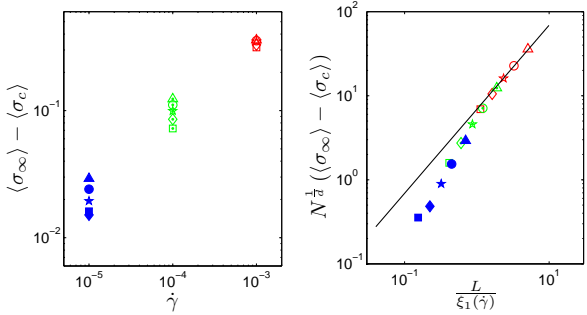}
\caption{Left panel: the average difference between the steady state stress $\sigma_\infty$ and the corresponding stress at the local minimum $\sigma_c$ for different system sizes as a function of $\dot\gamma$ in two dimensions. System sizes are $N=10000$ ($\triangle$) , $N=4096$ ($\bigcirc$) , $N=2116$ ($\bigstar$), $N=1024$ ($\lozenge$), $N=484$ ($\square$). Strain rates are  $\dot\gamma=10^{-5}$ (full symbols), $\dot\gamma=10^{-4}$ (dotted symbols) and $\dot\gamma=10^{-3}$ (empty symbols). Right panel: Data collapse by re-plotting the same data as shown. The continuous line has a slope of unity.}
\label{confstress}
\end{figure}

We expect that the mean drop in stress between the flow stress and the configurational stress should be determined by the ratio of the system size to the length scale $\xi_1$ of Eq. (\ref{xi1}). We thus propose a scaling function
\begin{equation}
\langle \sigma_\infty -\sigma_c \rangle \sim N^{\delta}  g_1 (L/\xi_1) \ . \label{g1}
\end{equation}
The dependence on $N^{\delta}$ is called for by the clear $N$ dependence seen in the left panel of Fig. \ref{confstress}.
The exponent $\delta$ and the asymptotics of the scaling function are determined by (i) requiring the loss
of the $N$ dependence when $\dot\gamma\to \infty$ and (ii) agreement with the Herschel-Bulkley law. We thus require
$g_1(x) \to x^{\hat\zeta}$ when $x\to \infty$, and write
\begin{equation}
\delta +\hat\zeta/d =0\ , \quad \hat\zeta/(1-\beta d) =\chi \ . \label{relations}
\end{equation}
The best data collapse is obtained with $\hat\zeta =1$;  the correctness of the
scaling ansatz is shown in the right panel of Fig. \ref{confstress} in which the data in the left panel
is collapsed to a single function using the rescaling proposed by Eq. (\ref{g1}).
The straight line in the right panel has a slope of unity to exemplify that $\hat\zeta=1$. As a consequence of this
result we find that $\delta=-1/d$ (i.e. 1/2 in 2D and 1/3 in 3D). Finally $\chi=1/(1-\beta d)$ which in 2D translates
to $\chi = 3/7$.

\begin{figure}
\centering
\includegraphics[scale = 0.40]{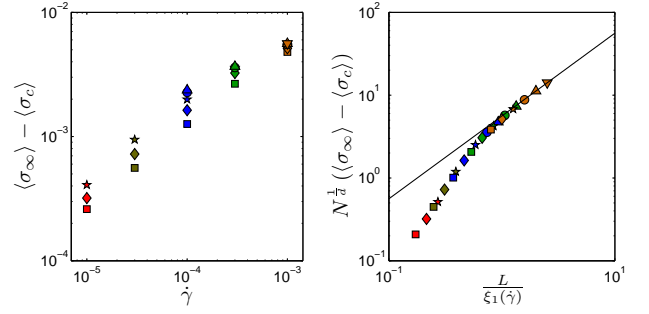}
\caption{Left panel: the average difference between the steady state stress $\sigma_\infty$ and the corresponding stress at the local minimum $\sigma_c$ for different system sizes as a function of $\dot\gamma$ in three dimensions. System sizes are $N=16000$, ($\bigtriangledown$), $N=8000$ ($\triangle$) , $N=4000$ ($\bigcirc$) , $N=2000$ ($\bigstar$), $N=1000$ ($\lozenge$), $N=512$ ($\square$). Right panel: Data collapse by re-plotting the same data as shown. The continuous line has a slope of unity.}
\label{confstress3D}
\end{figure}

These results continue to hold in three dimensions as well.
We have measured $\beta$ directly in 3D simulations and got the same number $\beta\approx -2/3$, cf. Fig. \ref{meanScaling}. Measuring the analog of Fig. \ref{confstress} in 3D we obtained the data shown in Fig. \ref{confstress3D}.
Indeed the result $\hat\zeta=1$ continues to hold, consistent with $\delta=-1/d$ and $\chi$ in 3D being
$\chi=1/3$.

\subsection{Cross over in the fluctuation spectra due to strain rate}
\label{cross2}

As in the case of temperature, also strain rate destroys the highly correlated plastic events that
are observed in AQS simulations. Again the relevant length scale should be $\xi_1$ of Eq. (\ref{xi1}), and
we construct a scaling function to describe the change in the nature of the fluctuations using this same length scale.
\begin{figure}
\centering
\includegraphics[scale = 0.40]{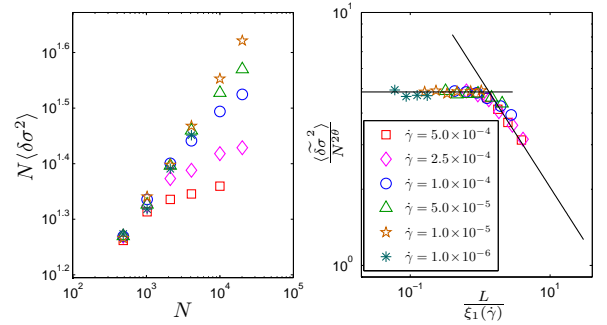}
\caption{Left panel: The stress variance in athermal simulations (multiplied by $N$) for different strain rates as a function of the system size, see inset in the right panel. The change in scaling upon increasing the strain rate
is obvious. Right panel: the data collapse in agreement with the scaling function $\tilde g_1$ of Eq. (\ref{tildeg})}
\label{crossgam}
\end{figure}
The data for the stress variance as a function of system size for different values of the strain rate
are shown in the left panel of Fig. \ref{crossgam}. The change in scaling with increasing $\dot\gamma$
is obvious. Here we propose a scaling function in the form $\sim N^{2\theta}\tilde g(x)$, or explicitly
\begin{equation}
\widetilde{\langle \delta \sigma ^2} \rangle  \sim \bar s^2 N^{2\theta}\tilde g(L/\xi_1(\dot\gamma)) \ . \label{tildeg}
\end{equation}
For $\dot\gamma\to 0$ (equivalently $\xi_1\to \infty$ or $x\to 0$) we should recover the AQS results, requiring
$\tilde g(x)\to$const. On the other hand when $\dot\gamma$ is large, or $x\to \infty$, we should get uncorrelated
fluctuations, $\widetilde{\langle \delta \sigma ^2\rangle} \to N^{-1}$. This requires $\tilde g(x) \to x^{\tilde \zeta}$ when $x \to \infty$
with $\tilde \zeta = 2(-2\theta -1)$. These asymptotics, the cross over at $x=1$, and the excellent data collapse are all seen in the right panel of Fig. \ref{crossgam}, adding full justification to estimate and the
ramifications of the existence of the the typical scale $\xi_1$.

\section{The Unifying Scaling Theory}
\label{scale}

In this section we provide a scaling theory that will unify the simulation results for all
temperature and strain rates.
Having two length scales at our disposal, we realize that the shortest of the two will be dominant at any given
 conditions $(T,\dot\gamma)$. We thus define $\xi$ according to :
\begin{equation}
\xi(T,\dot{\gamma}) = \left\{
\begin{array}{cc}
\xi_{1}\ , & T < T^* \\
\xi_2\ , & T > T^*
\end{array}\right.\ ,
\end{equation}
where $T^*$ is obtained by equating the two scales, i.e.
\begin{equation}
T^* \sim  \bar \epsilon \left[\left(\frac{\bar\epsilon}{\sigma_Y  \lambda^d}\right)\left(\frac{c}{  \lambda \dot\gamma}\right) \right]^{\beta d/(1-\beta d)} \ .
\end{equation}

 \begin{figure*}
\centering
\includegraphics[scale = 0.45]{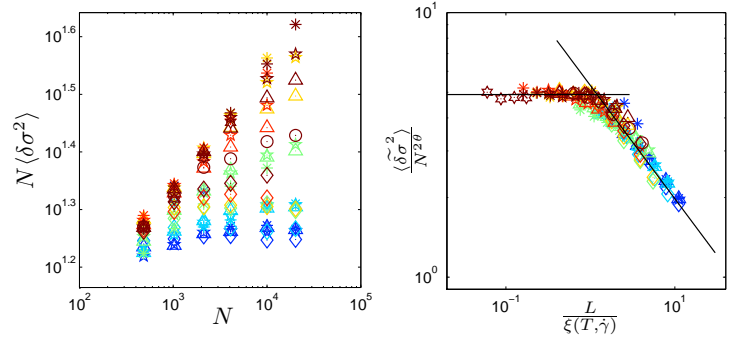}
\caption{Left panel: the measured stress fluctuations as a function of the system size for different temperatures and strain rates. Here the temperature ranges from $T=0$ to $T=0.15$ and the strain rate ranges from $\dot\gamma=1\times 10^{-6}$ to $\dot\gamma = 5\times 10^{-4}$. Right panel: data collapse as predicted by the scaling function Eq. (\ref{scalefun})}
\label{unifscale}
\end{figure*}

A very interesting and direct way of demonstrating the cross-over due to the combined thermal and finite strain-rate effects is provided by measurements
of the variance of the stress fluctuations as a function of the temperature, the strain rate and the system size. In Fig. \ref{unifscale} we display 2D measurements of this quantity which is obtained by averaging the square of the microscopic stress fluctuations in long stretches of elasto-plastic steady-states of the model described above at varying strain rates and temperatures as described in the figure legend.  The variance of the stress fluctuations decreases as a function of $N$, and in the left
panel we multiplied the variance by $N$ for better representation. Under quasi-static and athermal conditions the dependence is a power-law, cf. Eq.~(\ref{stressFlucsScaling}), with
$\theta\approx -0.4$ in the present model.  At higher temperatures and higher strain rate the data in Fig. \ref{unifscale} indicate a clear cross-over to normal intensive stress fluctuations in which
$\langle \delta \sigma^2\rangle \sim N^{-1}$.

As done before, to demonstrate unequivocally the cross-over that depends on both the temperature and the strain rate, we first need to separate the thermal from the mechanical contributions to $\langle \delta \sigma^2\rangle$ cf. Eq. (\ref{sigandT}). Using the scale $\xi$, we write the fluctuations in the form
\begin{equation}
\langle \delta \sigma^2 \rangle - \frac{\mu T}{V} \sim \bar s^2
N^{2\theta}{\cal F}\left(\sFrac{L}{\xi(T,\dot{\gamma})}\right)\ . \label{scalefun}
\end{equation}
The asymptotics of this function are again ${\cal F}(x)\to$ const for $x\to 0$ and ${\cal F}(x)\to x^{\tilde\zeta}$
for $x\to \infty$ as is born out in Fig. \ref{unifscale}.

\section{Concluding Remarks}
\label{conclusions}

We have shown in this paper that the mechanical response of amorphous solids to external strains
cannot be described with a theory that assumes a desert in which only thermal fluctuations are important
in triggering plastic events. Quite on the contrary, the problem abounds with rich and interesting
correlated fluctuations which are most spectacular in the AQS limit. This limit can be fully understood on the
basis of the mechanical instabilities that are seen as eigenvalues of the Hessian matrix going to zero. There
plastic events are highly correlated and appear in the form of system spanning avalanches. Increasing the temperature and strain rate has a great effect on these correlated plastic events, chopping them off and tending to make
them normal in the limit of high temperature and/or high strain rate. We explained that the mechanisms for
chopping off the correlations differ for temperature and strain rate, leading to a complicated cross-over
behavior when $T$ and $\dot \gamma$ increase. There are three typical regimes, the AQS regime $T<T_{\rm cross}$,
an intermediate regime $T_{\rm cross}<T<T_{\rm max}(\dot\gamma)$, and a thermal regime when $T>T_{\rm max}(\dot\gamma)$. The intermediate regime is most interesting with its gradual change in the statistical
physics of the system response and fluctuations. Interestingly, we demonstrated that scaling concepts are
of great importance in organizing the complex physics discovered and discussed above. All the effects of cross-over
could be captured with the help of judiciously chose scaling functions whose existence means a high degree
of predictability. By making measurements at some corner of the parameter space ($T$,$\dot\gamma$) we can
predict the correct results for any other point in this parameter space with the help of the scaling functions
presented above.

\acknowledgments
This work had been supported in part by the Israel Science Foundation and the Ministry of Science under the French-Israeli collaboration. We thank J. Chattoraj, A Lema$\^{\rm i}$tre and C. Caroli for sharing with us their results and ideas concerning the activation over strain dependent barriers prior to publication. These ideas have influenced our chapter IV B.

\end{document}